\documentclass[preprint,review,12pt]{elsarticle}                                
\usepackage[utf8]{inputenc}
\usepackage[T1]{fontenc}

\usepackage{amssymb}

\journal{Journal of Molecular Liquids}

\begin{document}

\begin{frontmatter}



\title{Are nucleation bubbles in a liquid all independent?}


\author{Jo\"el Puibasset\fnref{label1}}
\ead{puibasset@cnrs-orleans.fr}

\affiliation[label1]{organization={ICMN, CNRS, Universit\'e d'Orl\'eans},
            addressline={1b rue de la F\'erollerie, CS 40059}, 
            city={45071 Orl\'eans Cedex 02},
            country={France}}

\begin{abstract}
 The spontaneous formation of tiny bubbles in a liquid is at the root of the nucleation mechanism during the liquid-to-vapor transition of a metastable liquid. The smaller the bubbles the larger their probability to appear, and even for moderately metastable liquid, it is frequent to observe several tiny bubbles close to each other, suggesting that they are not all independent. It is shown that these spatially correlated bubbles should be seen as belonging to one single density depression of the liquid due to fluctuations (called LDR for Low Density Region) and should be counted as one event instead of several. This has a major impact on the characterization of the bubble density in a liquid, with consequences (i) for understanding liquid-to-vapor transitions which proceed through growing and merging of these correlated bubbles, and (ii) for free energy profile and barrier calculations with molecular simulation techniques which require to convert the calculated size distribution of the largest bubble into the size distribution of any bubble. Remarkably, the average number of LDRs in a given volume simply relates to the probability of not having bubbles in the liquid.

\end{abstract}

\begin{graphicalabstract}
\includegraphics[width=0.8\textwidth]{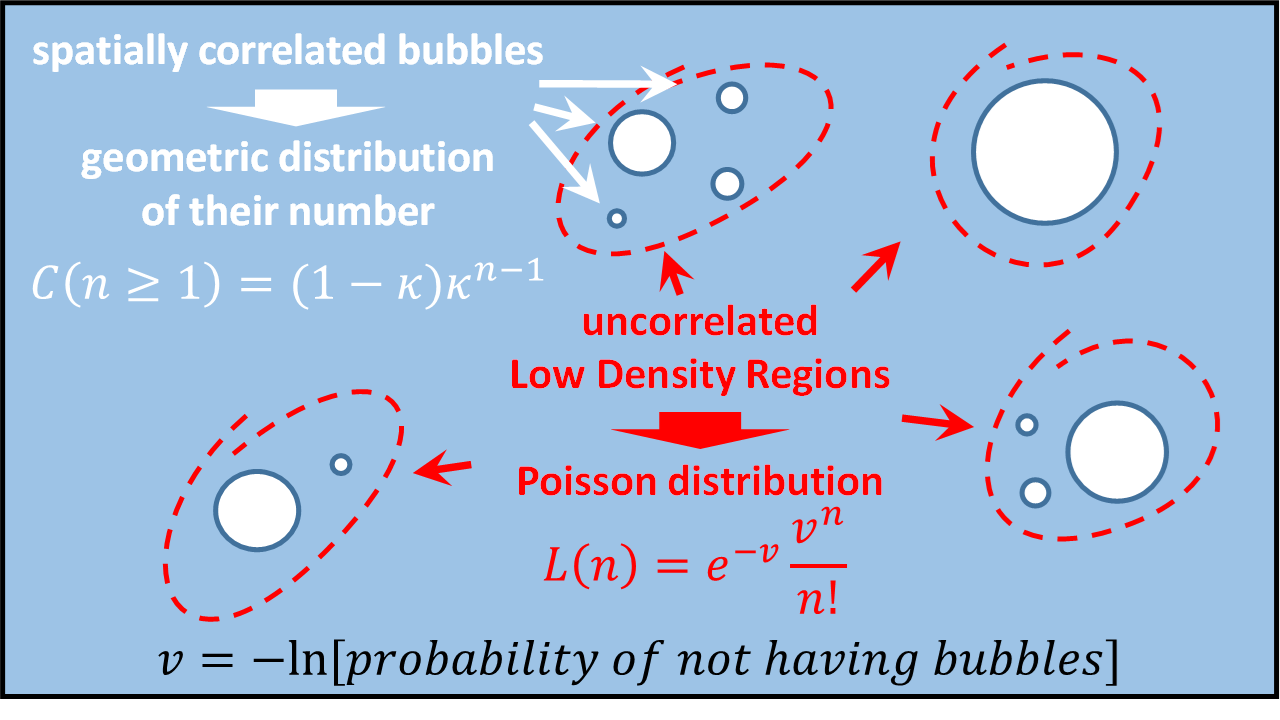}
\end{graphicalabstract}



\end{frontmatter}


\section{Introduction}

The structure of fluids is an important feature in relation with their thermodynamic properties, not only in terms of atomic or molecular pair correlations, but also in terms of voids or cavities between molecules \cite{RN2963, RN2859, RN2875, RN2984, RN2869, RN2961}. 
One can cite for example their importance to understand the solubility of large molecules which is related to the ability of the fluid to form cavities of sizes comparable to that of the solute \cite{RN2861, RN2871, RN2983}. 

In the context of phase transitions, in particular the liquid-to-vapor transition of a metastable superheated or stretched liquid (cavitation) \cite{RN2997, RN2760, RN2758, RN2715, RN2845, RN2972, RN2996, RN2991, RN3000, RN2989}, the classical nucleation theory also relies on the idea that spontaneous cavities can appear in the liquid \cite{RN548, RN512, RN2920}. 
As a matter of fact, such cavities are expected to appear spontaneously in any thermodynamic conditions, in metastable state or at equilibrium, the probability of occurrence being given by the Boltzmann factor \cite{RN2859, RN2913}
\begin{equation}
p(s) \propto e^{-W(s)/kT}\label{eq_boltz}
\end{equation}
where $W(s)$ is the free energy of formation of a cavity of size $s$, $k$ is Boltzmann's constant and $T$ the temperature. 
This relation is important in the context of molecular simulations, since, in principle, a simple measurement of the size of the spontaneous cavities in the liquid gives their relative free energies through the determination of $p(s)$.

Another important quantity is the number density of cavities in the liquid, regardless of their size. It appears at least in two important situations:

(i) In the context of nucleation, where the nucleation rate, as given by the classical nucleation theory, is proportional to the number density of initial germs \cite{RN548, RN512, RN2920}. It also raises the subsequent question of the identification of germs with the budding bubbles and more generally the bubble growth mechanisms \cite{RN2756, RN2968, RN2985, RN2770, RN598, RN2903, RN2766, RN2797}.

(ii) In the context of molecular simulations, the distribution of the number of bubbles $B(n)$ in a given liquid volume $V$ is also an important piece of information to determine the free energy profile from Eq.~(\ref{eq_boltz}). 
The reason is as follows: since it is impossible to observe spontaneous formation of large bubbles in molecular simulations due to small system sizes and short simulation runs, Frenkel and collaborators proposed to use biased Monte Carlo to sample bubbles up to the critical size \cite{RN2901, RN2909, RN1956, RN2911}.
However, the obtained size distribution is not $p(s)$ but the size distribution of the \emph{largest} bubble $p_L(s)$ which should not be used in Eq.~(\ref{eq_boltz}), in particular for small cavities \cite{RN2905, RN2893, RN2895, RN2907, RN2847, RN2801, RN2903}.
But it has been shown very recently that it is actually possible to transform $p_L(s)$ into $p(s)$ thanks to the distribution $B(n)$ \cite{RN2954}.
    
What is the general form for $B(n)$? Thirty years ago, it was proposed that the number $n$ of bubbles in a given volume follows the geometric law $\nu^n$ \cite{RN2913, RN2911, RN2966}, while more recently it was suggested that it should follow the Poisson law $\nu^n/n!$ \cite{RN2893}. 
Given the rapid increase of $n!$ it is important to propose arguments to decide between the two laws.
Let us denote $\nu$ the average number of bubbles in the volume. In the context of spontaneous cavity formation in a liquid that is not under extreme stress or superheat, $\nu$ is small, and therefore it corresponds to the probability to have one bubble. The probability to have $n$ independent bubbles is then argued to be proportional to $\nu^n$ \cite{RN2913, RN2911, RN2966}.
However, one has to take into account the fact that the $n$ bubbles appear in the \emph{same} volume and not in $n$ replicas. This is analogous to the decay problem which considers the probability to have $n$ decays in the \emph{same} interval of time. The consequence is the appearance of the factor $1/n!$, giving the Poisson law (see appendix).

\begin{figure}[]
\begin{centering}
\includegraphics[width=0.8\linewidth]{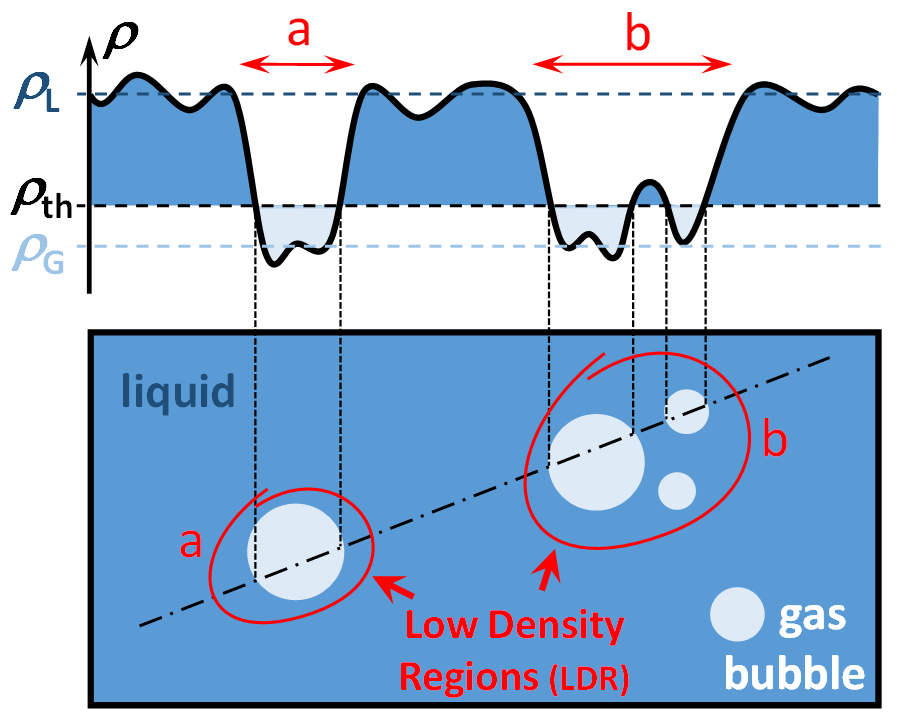}
\caption{\label{fig_liq_density} Schematic representation of the fluid density $\rho$ (upper black curve) along an arbitrary line passing through three bubbles in this example (dash-dotted line shown below), showing fluctuations around the average liquid density $\rho_{\rm L}$ (upper horizontal dashed line in dark-blue). Due to large localized fluctuations, some regions of space can exhibit a density lower than a given threshold $\rho_{\rm th}$ (black horizontal dashed line) defining it as a bubble of gas (e.g. the region denoted a). In some cases, depending on the choice of the threshold, the low density region (LDR) may appear as divided in several bubbles (e.g. region b). }
\end{centering}
\end{figure}

However, the spontaneous formation of tiny cavities is not so rare since the free energy cost is low. It is thus possible to have two or more bubbles in a small volume that could possibly interact if their distance is below the typical range of interactions in the liquid.
Furthermore, the underlying mechanism of bubble formation may produce correlations. For instance, Neimark and co-workers have shown that the growth of a bubble in a stretched liquid approaching its spinodal proceeds via the formation of multiple interstitial cavities that coalesce into a larger one \cite{RN598}.
Far from spinodal, the formation of bubbles is a consequence of localized density fluctuations in the liquid. As sketched in Fig.~\ref{fig_liq_density}, a bubble corresponds to a region of the liquid where the local density falls below a given threshold $\rho_{\rm th}$ between liquid and vapor densities (see for example the region denoted a). 
Depending on fluctuation correlations in the liquid, it is possible to have an extended region of low average density consisting of two or more neighboring bubbles separated by intervals where the density may locally exceed the threshold $\rho_{\rm th}$ (Fig.~\ref{fig_liq_density}, region b). Such bubbles are obviously not independent, and the associated number distribution is not expected to be Poissonian.

From a fundamental point of view, it is important to distinguish between these two situations corresponding to isolated versus correlated bubbles. For instance, in the context of nucleation, independent (isolated) bubbles can be considered as distinct nuclei, while closely correlated bubbles should be seen as one single nucleus (the growing bubbles close to each other are expected to coalesce rapidly) \cite{RN598}.
Therefore, in this case, the low density regions appear as a better concept to define the nuclei.
They will be denoted LDR for Low Density Region, to distinguish them from the cavities or bubbles.
It is important to note that, although the exact number of bubbles in a given LDR obviously depends on the threshold $\rho_{\rm th}$ and more generally on the procedure used to define the gas-like regions, the existence of this LDR is intrinsic and essentially independent of the procedure. 

The paper decomposes as follows. The first section presents the molecular liquid model used for illustration, the method used to map the volume into liquid-like and gas-like cells on a discrete grid, and the general procedure to analyze the correlations between the bubbles. 
The second section presents the numerical results, in particular the distribution of the number of bubbles $B(n)$ for two system sizes and two cluster algorithms to define the bubbles. 
Exploiting a correlation criterion between bubbles, the distributions $L(n)$ of the number of LDRs per configuration and $C(n)$ of the number of correlated bubbles per LDR, are established as a function of this criterion, showing an optimal choice. 
Then follows a discussion showing in particular that it is possible to decompose the distribution $B(n)$ of the number of bubbles in a given volume into a Poisson distribution of the number of LDRs (compatible with the hypothesis of independence) and a geometric distribution of the number of bubbles in a given LDR. 
It is shown in particular that the parameter associated with the Poisson distribution of LDRs is essentially independent of the algorithm used to define the bubbles. This allows in particular to define an intrinsic average number of LDRs relevant for nucleation.
The most striking result of the analysis is that this average number of LDRs can easily be deduced from molecular simulations by simply evaluating the probability of not having bubbles in a given volume of liquid.

\section{Methods and theory}
\emph{Molecular model:} The numerical model of the liquid is a Grand Canonical Monte Carlo treatment of generic spherical molecules interacting via the truncated and quadratically shifted Lennard-Jones potential with a cutoff radius equal to three atomic diameters \cite{RN154,RN51,RN1330}.
Note that the general features described in this paper are expected to be quite general and weakly dependent on the particular choice of interacting potentials. 
One has however to keep in mind that specific behaviors could emerge for systems that depart significantly from these assumptions, in particular those with long-range interactions (electrostatic), strongly anisotropic molecules, or many-body interactions (metals). 
It is also emphasized that the results are expected to be dependent on the statistical ensemble since it controls the density fluctuations, but no differences were observed between Monte Carlo and molecular dynamics runs.
All quantities will be expressed in reduced units built with the Lennard-Jones parameters $\epsilon$ and $\sigma$ associated to interaction intensity and molecular diameter. 

To enhance statistics and favor density fluctuations, we work at the reduced temperature $kT/\epsilon=1.0$ and at $\ln \left( z \sigma^3 \right) = -3.20$ where $z = e^{\mu /kT}/\Lambda^3 $ is the affinity, $\mu$ is the chemical potential and $\Lambda$ is the thermal de Broglie wavelength. In these conditions, the average reduced pressure is $p\sigma^3/\epsilon=0.026$, a value below the reduced saturating pressure $p_{\rm sat}\sigma^3/\epsilon=0.055$ at the working temperature, which places the system in a metastable state (superheated). 
Although the formation of tiny bubbles is slightly favored by the metastability, no transition was observed during the simulation runs: the metastability is too weak to allow the liquid to overcome the energy barrier.
No bias is used neither: the observed bubbles are all spontaneous. 
Two cubic system sizes are considered, of edges $L = 9\sigma$ and $18\sigma$ with periodic boundary conditions, containing an average number of molecules equal to 412 and 3300 respectively (see Table~\ref{table1}).

\begin{table}
\caption{System size $L/\sigma$, average number of molecules $<N_{\rm Mol}>$, bubble algorithm, average number of bubbles $\nu_{\rm b}$, probability of not having bubbles in the liquid $B(0)$, $\lambda_{\rm 0} = -\ln B(0)$, Poisson distribution parameter $\lambda_{\rm bb}$ and geometric distribution parameter $\kappa_{\rm bb}$ for the optimal bubble-bubble correlation length $l_{\rm bb} = 3\sigma$, and $\lambda$ and $\kappa$ values from fits with Eqs~\ref{EqnB0} and \ref{EqnBn} (see text for more details on the definitions). }
\begin{tabular}{c c c c c c c c c c}
\hline\hline
 $L/\sigma$ & $<N_{\rm Mol}>$ & algorithm & $\nu_{\rm b}$ & $B(0)$ & $\lambda_{\rm 0}$  & $\lambda_{\rm bb}$ & $\kappa_{\rm bb}$ &  $\lambda_{\rm fit}$ & $\kappa_{\rm fit}$ \\ [0.5ex]
\hline
 9.0   & 412  & (f)   &   1.3    & 0.443   &   0.81   &   -    &   -    &  1.37   &   0.44    \\
 9.0   & 412  & (fev) &   1.0    & 0.443   &   0.81   &   -    &   -    &  0.98   &   0.20    \\ 
 18.0  & 3300 & (f)   &   6.8    & 0.0155  &   4.17   &  4.15  &  0.47  &  4.11   &   0.40    \\
 18.0  & 3300 & (fev) &   5.3    & 0.0155  &   4.17   &  4.40  &  0.24  &  4.42   &   0.15    \\ [1ex]
\hline
\end{tabular}
\label{table1}
\end{table}

\emph{Bubble identification and characterization:} There are several methods to detect and characterize the bubbles appearing in the liquid, relying on an estimation of the local fluid density. For instance one can perform a running Gaussian convolution on the atomic positions. A less time consuming method consists in working on a discrete grid. Obviously, different methods will produce different results, but the main features of the fluid structure (like large deviations from the liquid density) should be detected by all methods. 
In this study we have chosen a four-steps algorithm which has proven to be very efficient, the \emph{M-method} \cite{RN1184,RN2847}:

(i) For each molecule in the system, one counts the number of neighbors closer than $1.625\sigma$ (first minimum of the radial distribution function of the liquid in our thermodynamic conditions). If this number is larger or equal to 6 (based on the analysis of liquid and vapor phases in equilibrium), the molecule is labeled as liquid-like, otherwise it is labeled as vapor-like.

(ii) The volume of the simulation box is divided into cubic cells (voxels) of edge $l_{\rm cell}=0.5\sigma$.

(iii) The grid cells in the vicinity of a liquid-like molecule (center-to-center distance $<1.625\sigma$) are marked as liquid, while the others are marked as vapor.

(iv) We then perform a cluster analysis on the vapor cells to define the bubbles. Two algorithms are used: the cells are considered to belong to the same bubble if they are connected through the faces only (f), or through the faces, edges or vertices (fev). The corresponding bubbles will be denoted (f)- and (fev)-bubbles.

\begin{figure}[]
\begin{centering}
\includegraphics[width=0.8\linewidth]{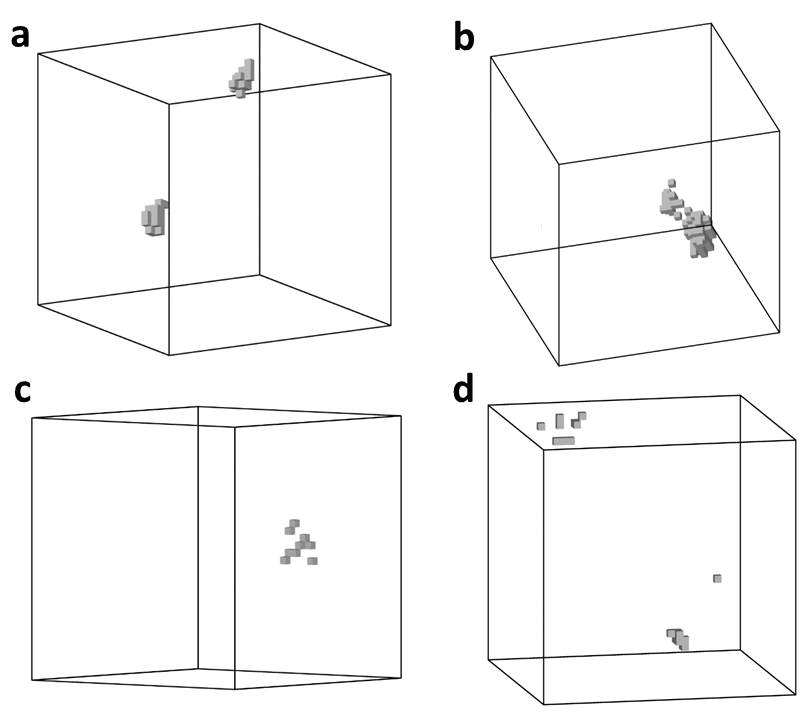}
\caption{\label{fig_bubbles} 3D representation of the result given by the M-method at step (iii) of the algorithm for four molecular configurations of the superheated Lennard-Jones liquid at $kT/\epsilon=1.0$ and $\ln z \sigma^3 = -3.20$ (see text). The grayed cells correspond to those marked as vapor, while the liquid cells are not shown.}
\end{centering}
\end{figure}

Figure~\ref{fig_bubbles} shows the result given by the previous algorithm at step (iii) for four molecular configurations.
The grayed cells in the figure correspond to those marked as vapor. A large variety of situations can be observed in these examples. 
In the first configuration (Fig.~\ref{fig_bubbles}a) one can see two well separated clusters of vapor cells unambiguously associated to two bubbles. In this case, the two cluster algorithms considered in step (iv) give the same result since all grayed cells are connected through their facets.

In the second configuration (Fig.~\ref{fig_bubbles}b) all vapor cells are localized in the same region of the simulation box. They form one single cluster if one includes vertex-sharing in the algorithm at step (iv) (giving one (fev)-bubble), while they separate into several bubbles if one considers only face-sharing connectivity (at least five (f)-bubbles can be seen). 
This is an example showing that the choice of a particular algorithm in step (iv) may give several bubbles in a situation where obviously only one nucleation is occurring since all vapor cells are closely connected. 

Figure~\ref{fig_bubbles}c gives an example where the two algorithms give several bubbles, because some clusters are separated from the others by at least one liquid cell. However, these clusters appear closely correlated since they are all localized in the same region of space compared to the simulation box volume, and most probably correspond to one single density fluctuation or transient  nucleation.

Finally, Fig.~\ref{fig_bubbles}d shows an even more complex situation where (i) the number of clusters depends on the algorithm, and (ii) the vapor cells are not all localized in the same region, suggesting the existence of several nucleation spots.  

These examples show the influence of the (f)- or (fev)-algorithms used in step (iv). Once the algorithm is specified, it is straightforward to identify the corresponding bubbles appearing in the liquid for each molecular configuration. 
It is then possible to monitor their number, size and position defined as that of their center-of-mass. 
The distribution of the number of bubbles can be cast into a histogram that can be accumulated over the configurations and normalized to one. 
The positions of the bubbles can also be accumulated over the whole simulation, and one can show that they are distributed uniformly over the simulation box.
Obviously, most of the quantities deriving from the bubbles are expected to depend on the cluster algorithm. It is however possible to extract intrinsic information on the LDRs, to be described now.

\emph{Low density regions (LDR) and their characterization:} To evidence the existence of LDRs and the correlations between the bubbles, it is interesting to calculate the average radial distribution function $g_{\rm bb}(r)$ between all bubbles in a configuration, where $r$ denotes their mutual distance. 
$g_{\rm bb}$ is build exactly as for the well-known $g(r)$ for atomic liquids, except that the positions are replaced by the center-of-mass of the bubbles, and the algorithm used to perform the average over the configurations is that used in the case of systems where the number of particles (bubbles in our case) varies (the self-averaging quantity is not $g_{\rm bb}$ but the two-point correlation density $\rho_{\rm bb}({\bf r_1},{\bf r_2})$ where ${\bf r_1}$ and ${\bf r_2}$ denote the positions in space of the two particles). 
As for a liquid, isotropy hypothesis is expected to hold despite the underlying discrete grid, which allows to perform an average over orientations. 

If the bubbles all nucleate independently, one expects a correlation function $g_{\rm bb}$ equal to 1 (ideal gas), while a peak would reveal correlations in positions. 
In that case, one may define a correlation length $l_{\rm bb}$ corresponding to this peak. Integration of $g_{\rm bb}-1$ times the bubble density over the peak would give the excess number of neighboring bubbles.
This correlation length $l_{\rm bb}$ gives a simple geometric criterion to define LDRs as clusters of correlated bubbles: two bubbles belong to the same LDR if they are within $l_{\rm bb}$. Two LDRs are then necessarily separated by a distance larger than $l_{\rm bb}$.

An important feature of LDRs which will be shown below is that they are intrinsic, in the sense that their existence and number is independent of the (f)- or (fev)-algorithm used to define the bubbles. 
Several questions regarding the LDRs then arise, necessitating the introduction of new distributions. For instance, what is the typical number of LDRs per configuration? And what is the typical number of bubbles per LDR?
More generally, we introduce the distribution $L(n)$ of the number of LDRs per configuration, and the distribution $C(n)$ of the number of correlated bubbles per LDR.
A general question to be addressed below is how the total distribution $B(n)$ relates to the distributions $L(n)$ and $C(n)$.

\section{Results}

\begin{figure}[]
\begin{centering}
\includegraphics[width=0.8\linewidth]{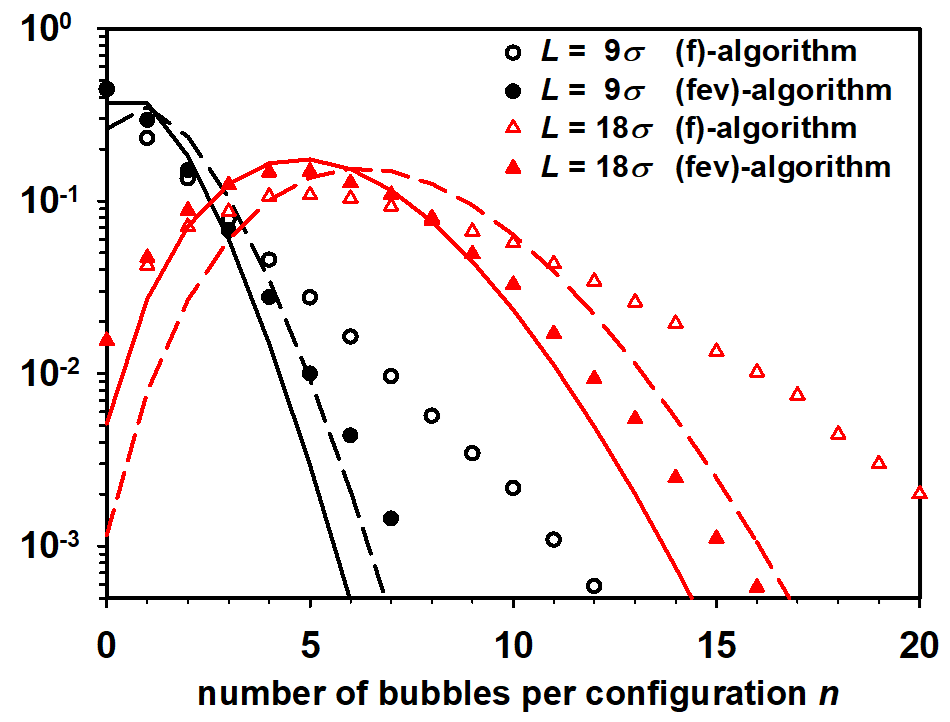}
\caption{\label{fig_distrib} Distributions $B(n)$ of the number of bubbles in the Lennard-Jones liquid at $kT/\epsilon=1.0$ and $\ln z \sigma^3 = -3.20$ for two systems sizes $L=9\sigma$ (black circles) and $L=18\sigma$ (red triangles). The bubbles are defined and build according to two cluster algorithms with criterion of face sharing (empty symbols, (f)-algorithm) or face-edge-vertex sharing (filled symbols, (fev)-algorithm). The dashed (respectively solid) lines are the Poisson distributions with the parameter equal to the corresponding average number of bubbles $\nu_{\rm b}$.}
\end{centering}
\end{figure}

\emph{Distribution of the number of bubbles $B(n)$:} Figure~\ref{fig_distrib} shows the distributions of the number of bubbles for the two system sizes $L=9\sigma$ and $18\sigma$ and the two (f)- and (fev)-algorithms used to define the bubbles. 

For the smallest system, the largest probability is for zero bubbles in the configuration. The probability then decreases monotonically with essentially an exponential decay above few bubbles for both algorithms.

\begin{figure}[]
\begin{centering}
\includegraphics[width=0.8\linewidth]{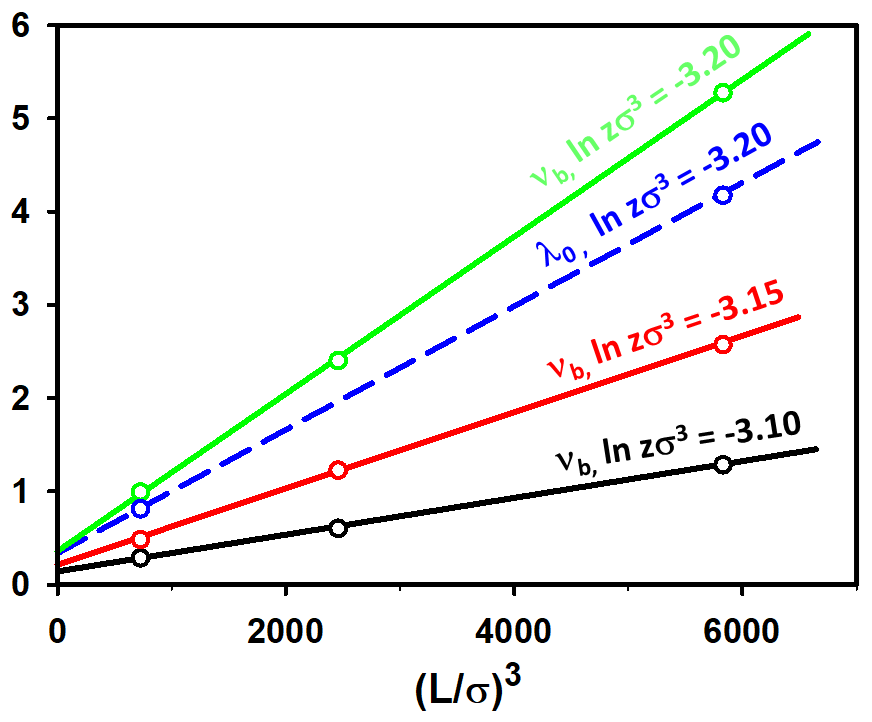}
\caption{\label{fig_prop} Symbols with solid lines: average number of (fev)-bubbles $\nu_{\rm b}$ as a function of the reduced system volume $(L/\sigma)^3$ at $kT/\epsilon=1.0$ and for three affinities $z$ given in the figure. The three system sizes are $L/\sigma=$ 9, 13.5 and 18. Symbols with the dashed line: evolution of $\lambda_{\rm 0}$ (see text) for $\ln z \sigma^3 = -3.20$ and for two system sizes. Lines are guide to the eye.}
\end{centering}
\end{figure}

For the largest system, the distributions exhibit a maximum: it is more probable to observe several bubbles instead of 0 or 1. This is expected since the average number of bubbles $\nu_{\rm b}$ increases with the system size, as shown in Fig.~\ref{fig_prop} where the average number of (fev)-bubbles is plotted versus the system size for three different affinities. Note however that one does not observe a strict proportionality with the volume. A small offset can be seen for all affinities, that can be attributed to finite size effects and possible correlations through the periodic boundary conditions for small systems. 

The distributions $B(n)$ are also sensitive to the cluster algorithm. Including edges and vertices produces a distribution which is shifted to the left (less bubbles) compared to that obtained with faces only, since two (f)-bubbles connected by a corner or an edge are considered as one single (fev)-bubble (see for example Fig.~\ref{fig_bubbles}b).

An important feature is that the first bin $B(0)$ is independent of the bubble algorithm (Fig~\ref{fig_distrib}). It corresponds to the probability that all cells of the volume are in the liquid state. The numerical values are reported in Table~\ref{table1}.  

Is there a general law to describe these distributions? If bubbles are supposed to be independent (like in Fig.~\ref{fig_bubbles}a exhibiting two well-separated bubbles), the distribution of their number is expected to be Poissonian (see Appendix), with a parameter equal to the average number of bubbles $\nu_{\rm b}$. Figure~\ref{fig_distrib} shows the normalized Poisson distributions $e^{-\nu_{\rm b}}\nu_{\rm b}^n/n!$ corresponding to the two system sizes and the two cluster algorithms (see Table~\ref{table1}). 
As can be seen, a fairly good agreement is found between Poisson distributions and data for the largest system $L=18\sigma$ and the (fev)-algorithm. 
However, noticeable discrepancies can be noticed: first, for the same (fev)-algorithm but a smaller system size $L=9\sigma$, the disagreement is noticeable, especially for large numbers of bubbles. The Poisson law significantly underestimates the probability to observe more than five bubbles in the system. 
Furthermore, for the (f)-algorithm, the disagreement is significant for both system sizes and for the whole range of the number of bubbles, which is particularly evident for the largest system. As a general trend one observes that the probability to observe small or large number of simultaneous bubbles is larger than predicted by the Poisson law with its parameter equal to the average number of bubbles, while the probability to observe a number of bubbles close to its typical or average value (around its maximum) is overestimated by the Poisson law. These observations clearly indicate that the detected bubbles are most probably not all independent, as suggested by the qualitative analysis of configurations (Fig.~\ref{fig_bubbles}). This is particularly true for (f)-bubbles, but also for (fev)-bubbles. 

\begin{figure}[]
\begin{centering}
\includegraphics[width=0.8\linewidth]{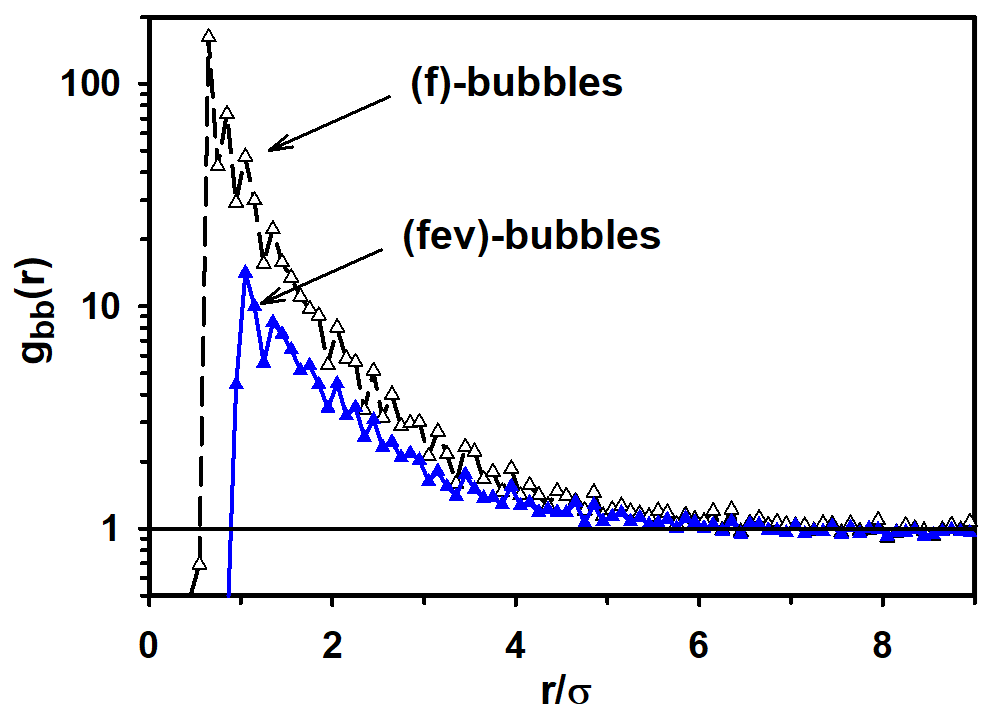}
\caption{\label{fig_gr} Radial distribution function $g_{\rm bb}(r)$ of the positions of the center-of-mass of the bubbles in the Lennard-Jones liquid at $kT/\epsilon=1.0$ and $\ln z \sigma^3 = -3.20$ for $L=18\sigma$ and for the (f)-algorithm (empty symbols) and the (fev)-algorithm (filled symbols).}
\end{centering}
\end{figure}

\emph{Correlations between bubbles:} Clearly, the independence assumption is not fulfilled in general, in particular for the (f)-algorithm, which obviously overestimates the number of bubbles in the liquid (Fig.~\ref{fig_bubbles}b). Despite a better agreement for the (fev)-algorithm, visual inspection of configurations shows that some bubbles may appear as not so well defined structures, with several disconnected sub-bubbles in the same region of space (see for instance Figs.~\ref{fig_bubbles}c and d).
For a quantitative evaluation of the correlations between the bubbles, the radial distribution function $g_{\rm bb}(r)$ has been calculated and reported in Fig.~\ref{fig_gr} for the largest system and the two algorithms. 

The rapid oscillations are due to the discretization of the underlying grid used to map the simulation box and should be ignored. 
As can be seen, $g_{\rm bb}(r)$ is zero below the shortest possible distance between unconnected vapor cells ($l_{\rm cell}\sqrt{2} = 0.5\sigma\sqrt{2}$ for (f)-bubbles and $2l_{\rm cell} = \sigma$ for (fev)-bubbles), and reaches 1 at large distances, as expected for uncorrelated positions. 
At intermediate distances, a strong correlation appears for (f)-bubbles, the height of the peak (roughly 100) being amplified by the low density of bubbles in the simulation box.
The peak then rapidly decreases (logarithmic scale): $g_{\rm bb} \simeq 10$ at $r = 1.8\sigma$ and 3 at $r = 3\sigma$. At $r=6\sigma$, $g_{\rm bb}$ has reached 1, meaning that correlations have disappeared. 
Interestingly, for (fev)-bubbles, the correlation function is significantly attenuated, which could be anticipated from the fact that the associated number distribution follows more closely a Poisson distribution.

Integration of $g_{\rm bb} - 1$ times the density of bubbles gives the total number of bubbles in excess around any given bubble, the excess being understood with respect to the situation where the positions are uncorrelated.
This quantity reaches 1.4 for (f)-bubbles and 0.4 for (fev)-bubbles, meaning that, in practice, only 1.4 (respectively 0.4) bubbles are found in the vicinity of any given bubble where the correlation function contributes the most.
Here again, the low excess found for the (fev)-bubbles suggests that they are less correlated. 

At odds with the usual radial distribution functions found in dense liquids, one does not observe a clear end to the correlation peak (associated to a minimum for instance). It is therefore difficult to give a precise value for the correlation length except that it is most probably in the interval $2\sigma \leq l_{\rm bb} \leq 6\sigma$.

\emph{Distribution of the number of LDRs $L(n)$:} The correlation length $l_{\rm bb}$ can be used to define the LDRs as ``clusters of correlated bubbles'', \textit{i.e.} two bubbles belong to the same LDR if their mutual distance is lower than $l_{\rm bb}$.
The configurations have been reanalyzed in terms of LDRs, for different values of $l_{\rm bb}/\sigma$ between 1 and 5, \textit{i.e.} in a broad range above the correlation peak of $g_{\rm bb}$.
Two quantities have been monitored: the distribution $L(n)$ of the number of LDRs in the configurations, and the distribution $C(n)$ of the number of bubbles per LDR. 

\begin{figure}[]
\begin{centering}
\includegraphics[width=0.8\linewidth]{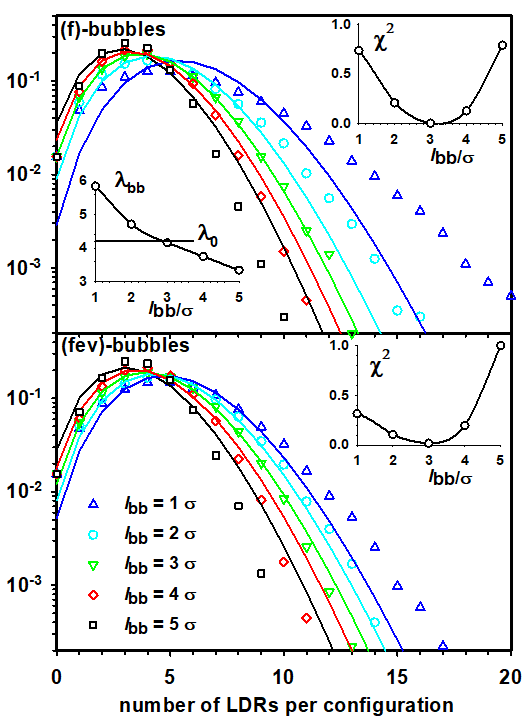}
\caption{\label{fig_nbLDR} Symbols: distributions $L(n)$ of the number of LDRs per configuration for the Lennard-Jones liquid at $kT/\epsilon=1.0$, $\ln z \sigma^3 = -3.20$, for $L=18\sigma$ and for different correlation lengths $l_{\rm bb}$ given in the figure. The lines are the corresponding Poisson distributions with parameters equal to the average number of LDRs $\lambda_{\rm bb}$. Upper panel: (f)-bubbles; lower panel (fev)-bubbles. The insets give the evolution with $l_{\rm bb}$ of $\lambda_{\rm bb}$ (for (f)-bubbles) and the goodness of fit $\chi^2$ (for (f)- and (fev)-bubbles).}
\end{centering}
\end{figure}

The results for $L(n)$ are given in Fig.~\ref{fig_nbLDR} for $L=18\sigma$ and the (f)- and (fev)-bubbles.
As expected, the distributions depend on the choice of the correlation length $l_{\rm bb}$, with a stronger effect for (f)-bubbles. By definition, for $l_{\rm bb}=0$ the distribution $L(n)$ identifies with that for all bubbles $B(n)$ given in Fig.~\ref{fig_distrib}.
Increasing $l_{\rm bb}$ induces a decrease in the probability to find a large number of LDRs, which mechanically induces an increase in the probability to find smaller numbers of LDRs. 
The variations with $l_{\rm bb}$ are qualitatively the same for (f)- and (fev)-bubbles except that for (fev)-bubbles $l_{\rm bb}=1\sigma$ gives essentially the same result as for $l_{\rm bb}=0$, and variations with $l_{\rm bb}$ are less pronounced.
It is remarkable that for large $l_{\rm bb}\geq 4$ the distributions found for (f)- and (fev)- bubbles tend to become identical.
An important feature is that the first bin $L(0)$ corresponding to the probability to find zero LDR is independent of $l_{\rm bb}$ since it is given by the probability to find zero bubble: $L(0) = B(0)$.

For all $l_{\rm bb}$, the corresponding Poisson distributions calculated with their parameter taken equal to the average number of LDRs (denoted $\lambda_{\rm bb}$) are shown as lines in Fig.~\ref{fig_nbLDR}. A goodness of fit $\chi^2$ is calculated which measures the departure of the data from the Poisson law (see insets in Fig.~\ref{fig_nbLDR}). 
As can be seen, an optimal value ($\chi^2$ approaching zero) can be found, which corresponds to $l_{\rm bb} \simeq 3\sigma$. The corresponding $\lambda_{\rm bb}$ values are 4.15 and 4.40 for the (f)- and (fev)-bubbles respectively (see inset giving $\lambda_{\rm bb}$ versus $l_{\rm bb}$ in Fig.~\ref{fig_nbLDR} and Table~\ref{table1}).
It is remarkable that the two (f)- and (fev)-algorithms finally give essentially the same value for the optimal correlation length.
This is an important argument in favor of the intrinsic nature of LDRs.

Let us now focus on a second way to determine the optimal $\lambda_{\rm bb}$. We already noticed that the first bin $L(0)$ has a constant value independent of $l_{\rm bb}$ ($L(0) = B(0) = 0.0155$, for $L = 18 \sigma$). Since, by definition of the Poisson law, the first bin is given by $e^{-\lambda_{\rm bb}}$, the optimal choice for the Poisson parameter is also expected to be $\lambda_{\rm bb}^{\rm optimal} = \lambda_{\rm 0} = -\ln L(0) = 4.17$. This value is reported as a horizontal line in the inset of Fig.~\ref{fig_nbLDR}. As can be seen, this value is in close agreement with $\lambda_{\rm bb} = 4.15$ found previously for $l_{\rm bb} \simeq 3 \sigma$ for (f)-bubbles (4.40 for (fev)-bubbles).  
The consistency of the analysis is also supported by the fact that the dependence of $\lambda_{\rm 0}$ with the system size closely follows the linear behavior observed for the total number of bubbles (see Fig.~\ref{fig_prop}).  

\begin{figure}[]
\begin{centering}
\includegraphics[width=0.8\linewidth]{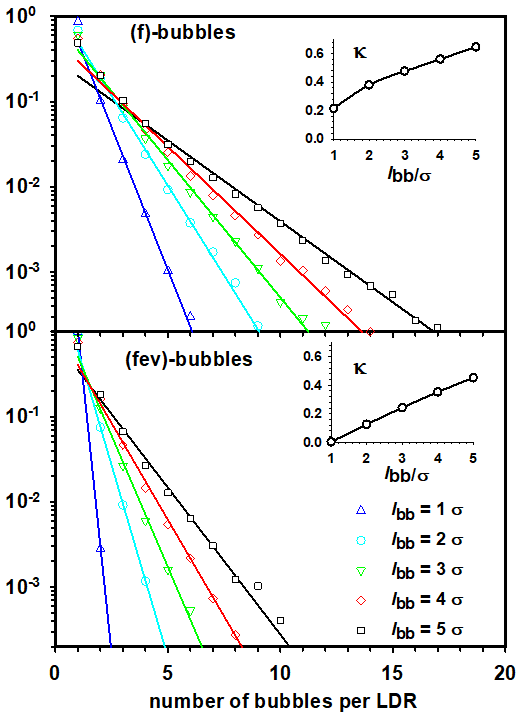}
\caption{\label{fig_bub_perLDR} Symbols: distributions $C(n)$ of the number of correlated bubbles per LDR for the Lennard-Jones liquid at $kT/\epsilon=1.0$, $\ln z \sigma^3 = -3.20$, for $L=18\sigma$ and for different correlation lengths $l_{\rm bb}$ given in the figure. The lines are guides to the eye corresponding to geometric distributions $\kappa^n$. Upper panel: (f)-bubbles; lower panel (fev)-bubbles. The insets give the evolution of $\kappa$ with $l_{\rm bb}$.}
\end{centering}
\end{figure}

\emph{Distribution of the number of bubbles per LDR $C(n)$:} Figure~\ref{fig_bub_perLDR} gives the distribution of the number of correlated bubbles in each LDR, averaged over all LDRs and configurations.
Note that these distributions are defined only for non-zero values $n \geq 1$.
As for $L(n)$, these distributions depend on the choice of $l_{\rm bb}$ and the bubble algorithm.  
Quite remarkably, except for the very first points, the general trend is a geometric law $\kappa^n$, as materialized by the straight lines in semi-log scale on Fig.~\ref{fig_bub_perLDR}. The slopes give the $\kappa$ values and its evolution with $l_{\rm bb}$ is given in the insets.
As can be seen, the distributions for the (fev)-bubbles decrease much faster than those for the (f)-bubbles as revealed by the lower $\kappa$ values. This is in agreement with the fact that the (f)-algorithm overestimates the number of bubbles per LDR. 
The obtained $\kappa_{\rm bb}$ values for the optimal choice $l_{\rm bb} = 3\sigma$ are reported in Table~\ref{table1}.

\section{Discussion}
\emph{A general model for $B(n)$:} The previous analysis in terms of LDRs can be considerably simplified by determining a general form for the distribution $B(n)$ based on the previous results. The optimal choice for $l_{\rm bb}$ gives a Poisson distribution for $L(n)$ compatible with the independence between the LDRs. 
We have also seen that $C(n)$ essentially follows a geometric distribution. Although a slightly more refined model could be used, it would have the inconvenience to introduce more free parameters.

It is therefore proposed to write $L(n)=e^{-\lambda}\lambda^n/n!$ where $\lambda$ is the first free parameter of the model and corresponds to the average number of LDRs per configuration. We also write $C(n)\propto \kappa^n$ for $n\geq 1$ with one single parameter. 
We use the convention $C(0)=0$ since by definition an LDR necessarily contains at least one bubble. 
Normalizing to one imposes $C(n\geq1)=(1-\kappa)\kappa^{n-1}$.

With this simple model, the distribution $B(n)$ of the number of bubbles per configuration can be easily deduced.
$B(0)$ being the probability to have zero bubble, one necessarily has zero LDR and thus 
\begin{equation}
B(0)= L(0) = e^{-\lambda}.
\label{EqnB0}
\end{equation}
To have exactly one bubble, it is necessary to have exactly one LDR containing one bubble: 
\begin{equation}
B(1) = L(1) C(1) = e^{-\lambda}\lambda (1-\kappa).
\end{equation}
For two bubbles, one can either have one LDR containing two bubbles or two LDRs each containing exactly one bubble: 
\begin{equation}
B(2) = L(1) C(2) + L(2) C(1)^2 = e^{-\lambda}\lambda (1-\kappa)\lbrace \kappa + \lambda (1-\kappa)/2 \rbrace.
\end{equation}
For three bubbles there are four terms: one can either have one LDR containing three bubbles, or two LDRs with one bubble in the first and two bubbles in the second, or the converse, or three LDRs each containing one bubble:
\begin{eqnarray}
B(3) = L(1) C(3) + L(2) (C(1)C(2)+C(2)C(1)) + L(3) C(1)^3 \nonumber \\
= e^{-\lambda} \lambda (1-\kappa)\lbrace \kappa^2 + \lambda \kappa(1-\kappa) + \lambda^2 (1-\kappa)^2/3! \rbrace.
\end{eqnarray}
The generic expression for $B(n \geq 1)$ decomposes into a sum over the number $k$ of LDRs, where each term comprises all combinations of $n$ bubbles distributed into $k$ LDRs:
\begin{equation}
B(n \geq 1) = \sum_{k=1}^{n}L(k)\sum_{i_1,...,i_k\in \lbrace 1...n\rbrace \atop i_1+...+i_k=n } C(i_1)C(i_2)...C(i_k).
\end{equation}
In the case where $C(i)$ follows a geometric law, all the terms in the second sum are identical to $(1-\kappa)^k\kappa^{n-k}$, and the number of terms equals the number of ways to place $k-1$ graduations on a discrete (integer) segment of length $n-1$, which equals $n-1 \choose k-1$. One then gets (note that the sum now runs from 0 to $n-1$): 
\begin{equation}
B(n \geq 1) = e^{-\lambda} \lambda (1-\kappa) \sum_{i=0}^{n-1}\frac{1}{(i+1)!}{n-1 \choose i}\lambda^i (1-\kappa)^i \kappa^{n-1-i}.
\label{EqnBn}
\end{equation}

\begin{figure}[]
\begin{centering}
\includegraphics[width=0.8\linewidth]{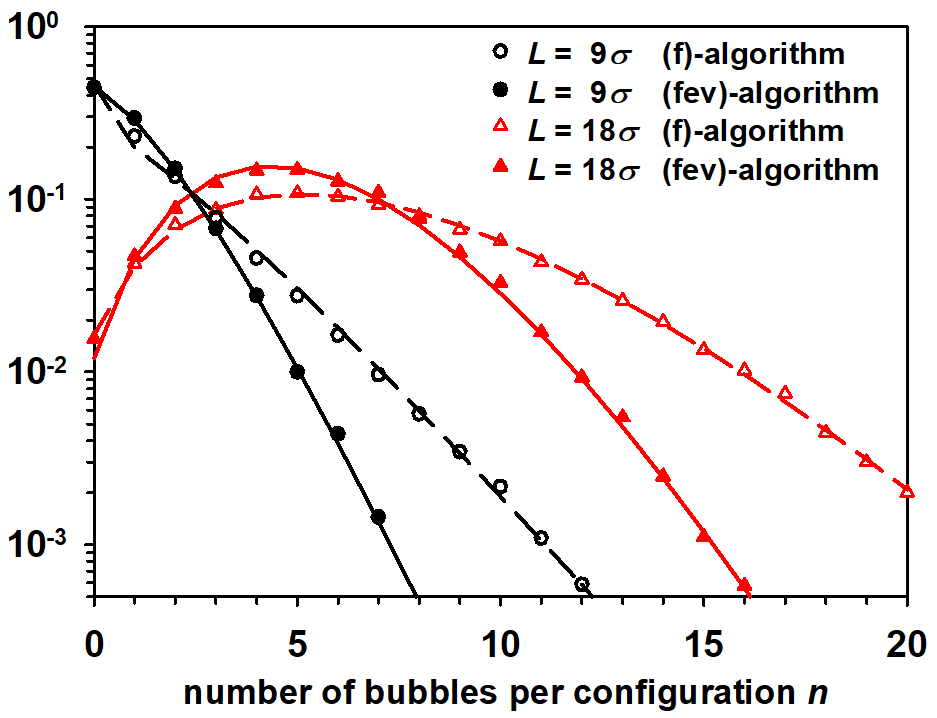}
\caption{\label{fig_fits} Same as Fig.~\ref{fig_distrib} except that the solid and dashed lines are now the best fits obtained with Eqs~\ref{EqnB0} and \ref{EqnBn} and the parameters given in Table~\ref{table1}.}
\end{centering}
\end{figure}

\emph{Application to data analysis:} The obtained expression is quite simple and can easily be used along with Eq~\ref{EqnB0} to fit the data with the two free parameters $\lambda$ and $\kappa$. 
The result of the fits are given in Fig.~\ref{fig_fits} for the two system sizes and the corresponding values of the free parameters are given in Table~\ref{table1}. 
As can be seen, the proposed model gives a remarkable agreement with the data, for (f)- and (fev)-bubbles as well. 
For the largest system $L=18\sigma$ the $\lambda$ parameters are 4.11 and 4.42 for (f)- and (fev)-bubbles respectively, \textit{i.e.} identical within 7\%, showing that the model gives essentially the same average number of LDRs for both algorithms. 
The difference is larger for the smallest system, most probably due to larger uncertainties (lower statistics). 
On the other hand, the $\kappa$ parameters are significantly different between the two algorithms. 
It is larger for (f)-bubbles, reflecting the fact that LDRs are split in a larger number of bubbles with the (f)-algorithm. It is actually not surprising to observe this dependence with the algorithm, since we have already seen that the (f)-algorithm produces more correlated bubbles. 
These observations (small difference for $\lambda$ versus a larger one for $\kappa$) suggest that the proposed model (Eqs~\ref{EqnB0} and \ref{EqnBn}) correctly absorbs the differences between the two algorithms in the $\kappa$ parameter, allowing to extract the intrinsic number of LDRs $\lambda$.

How does it compare with our previous analysis in terms of correlation length between bubbles? As can be seen in Table~\ref{table1} the results of the fit are in close agreement (within 1\%) with the $\lambda_{\rm bb}$ values that were extracted from the analysis consisting in optimizing the Poisson fit by varying $l_{\rm bb}$. The agreement with $\kappa_{\rm bb}$ is less good, the difference being 15\% for (f)-bubbles and reaching 45\% for (fev)-bubbles. The main reason is that the best fit with Eqs~\ref{EqnB0} and \ref{EqnBn} imposes a geometric law for all number of bubbles $\geq 1$, while the $\kappa$ values deduced from the slopes in the semi-log representations in Fig~\ref{fig_bub_perLDR} ignore the very first points, and are thus more reliable. Note that it is possible to improve the fits by replacing the single-parameter distribution $C(n\geq 1) \propto \kappa^n$ by the two-parameter law $C(1) = \alpha$ and $C(n\geq 2) \propto \kappa^n$.

\section{Conclusion}

Using molecular simulations of a simple van der Waals fluid, we have shown that the liquid phase exhibits spontaneous formation of tiny bubbles. It is emphasized that the precise definition of these tiny bubbles is not unique and sensitive to the algorithm used. Inspection of the distribution of the number of bubbles $B(n)$ in a given volume reveals that it does not follow the Poisson law $e^{-\lambda}\lambda^n/n!$ expected for uncorrelated events. 
We have shown that the bubbles are actually spatially correlated, with a typical length scale of order $3\sigma$, \textit{i.e.} in the range of the intermolecular interactions. 
Such correlated bubbles are expected to coalesce in the case of a growing nucleation event (in a metastable liquid), and should be seen as one single event instead of several bubbles. 
These correlated bubbles are interpreted as belonging to the same fluctuation in a region of space where the density is locally gas-like (called LDR for Low Density Region). 
This refined analysis in terms of LDRs allows to determine their average number in the simulation box, which appears to be independent of the algorithm used to define the bubbles, emphasizing their intrinsic nature. 
Their distribution is shown to follow the Poisson law, as expected for independent events. 
On the other hand, the simulations show that the distribution $C(n)$ of correlated bubbles in a single LDR is algorithm-dependent. It is however noticed that $C(n)$ essentially follows a geometric distribution for $n$ larger than few bubbles. 

Conversely, taking these observations as an assumption, it is possible to obtain a general expression for the total distribution $B(n)$ with only two free parameters (Eqs~\ref{EqnB0} and \ref{EqnBn}) that can be used to fit the data. 
The parameter associated to $C(n)$ is shown to be dependent on the algorithm used to define the bubbles, while the parameter $\lambda$ associated to the Poisson distribution of the uncorrelated LDRs is independent of the algorithm used to define the bubbles, which supports the idea that the LDRs are intrinsic. The obtained values are in good agreement with the previous analysis in terms of correlations between bubbles.  

A remarkable result is that the $\lambda$ parameter associated to the number of LDRs can easily be extracted from the data without need to determine the correlation length between bubbles nor to fit the data with (Eqs~\ref{EqnBn}). Thanks to Eq~\ref{EqnB0}, the relation $\lambda = -\ln B(0)$ immediately gives the desired average number of LDRs in the liquid. Note that the determination of $B(0)$, which is the probability of not having bubbles in the liquid, does not even require to use the (f)- or (fev)-algorithm to define the bubbles, since this probability is equivalent to having zero gas-like cell in the system, an information that can be determined at step (iii) of the M-method. This is particularly useful for the rapid determination of $\lambda$ entering the transformation of the distribution of the largest bubble given by biased simulations into the distribution of all bubbles entering Eqs~\ref{eq_boltz} \cite{RN2954}.

\section*{Declaration of interests}
The author declares to have no known competing financial interests or personal relationships that could have appeared to influence the work reported in this paper.

\section*{Acknowledgments}
The author acknowledges fruitful discussions with P. Porion, E. Rolley, P. Spathis and P.E. Wolf, and the financial support of Agence Nationale de la Recherche through the project CavConf, ANR-17-CE30-0002.

\section*{Data Availability Statement}
The data that support the findings of this study are available from the corresponding author upon reasonable request.

\appendix

\section{Distribution of the number of independent bubbles}
Let us consider a fixed volume $V$ of liquid with a uniform probability of having a nucleation (\textit{i.e.} a bubble) at any position. The volume can be divided in a large number $N$ of small cells of volume $dV = V/N$. We will suppose $dV$ infinitesimal ($dV \rightarrow 0$, $N \rightarrow \infty$). The probability to have a nucleation in $dV$ is also an infinitesimal, denoted $\tau dV$. The probability to have two nucleations or more in a cell is of order $dV^2$ and can be neglected. Therefore, the average number of nucleations in $dV$ equals $\tau dV$, and, the probability being uniform, the average number of nucleations in $V = N dV$ is given by $\nu = \tau V$. In the following, we suppose that the cells are independent, \textit{i.e.} the probability that a nucleation occurs in a cell is independent of the other cells.    

We are now in position to calculate the probability $B(n)$ to have $n$ independent bubbles in the volume $V$. This situation is obtained for any configuration where nucleation occurs in $n$ cells (probability $(\tau dV)^n$), while the other $N-n$ cells remain liquid (probability $(1-\tau dV)^{N-n}$), the total number of possible choice for the $n$ cells being $N \choose n$. Therefore, $B(n)$ is given by the limit:    
\begin{equation}
B(n) = \lim\limits_{N \rightarrow \infty} {N \choose n} \left[ \tau \frac{V}{N}\right]^n \left[ 1-\tau \frac{V}{N}\right]^{N-n} = e^{-\nu} \frac{\nu^n}{n!}
\end{equation}
which is the Poisson distribution.

\bibliographystyle{elsarticle-num} 
\bibliography{manuscript_bubbles_LDR}

\begin{thebibliography}{10}
\expandafter\ifx\csname url\endcsname\relax
  \def\url#1{\texttt{#1}}\fi
\expandafter\ifx\csname urlprefix\endcsname\relax\def\urlprefix{URL }\fi
\expandafter\ifx\csname href\endcsname\relax
  \def\href#1#2{#2} \def\path#1{#1}\fi

\bibitem{RN2963}
J.~G. Kirkwood, F.~P. Buff, The statistical mechanical theory of solutions.
  {I}, The Journal of Chemical Physics 19~(6) (1951) 774--777.
\newblock \href {https://doi.org/10.1063/1.1748352}
  {\path{doi:10.1063/1.1748352}}.

\bibitem{RN2859}
H.~Reiss, H.~L. Frisch, J.~L. Lebowitz, Statistical mechanics of rigid spheres,
  The Journal of Chemical Physics 31~(2) (1959) 369--380.
\newblock \href {https://doi.org/10.1063/1.1730361}
  {\path{doi:10.1063/1.1730361}}.

\bibitem{RN2875}
E.~Helfand, H.~Reiss, H.~L. Frisch, J.~L. Lebowitz, Scaled particle theory of
  fluids, The Journal of Chemical Physics 33~(5) (1960) 1379--1385.
\newblock \href {https://doi.org/10.1063/1.1731417}
  {\path{doi:10.1063/1.1731417}}.

\bibitem{RN2984}
D.~S. Corti, P.~G. Debenedetti, S.~Sastry, F.~H. Stillinger, Constraints,
  metastability, and inherent structures in liquids, Physical Review E 55~(5)
  (1997) 5522--5534.
\newblock \href {https://doi.org/10.1103/PhysRevE.55.5522}
  {\path{doi:10.1103/PhysRevE.55.5522}}.

\bibitem{RN2869}
P.~J. in~‘t Veld, M.~T. Stone, T.~M. Truskett, I.~C. Sanchez, Liquid
  structure via cavity size distributions, The Journal of Physical Chemistry B
  104~(50) (2000) 12028--12034.
\newblock \href {https://doi.org/10.1021/jp001934c}
  {\path{doi:10.1021/jp001934c}}.

\bibitem{RN2961}
J.~M. Simon, P.~Krüger, S.~K. Schnell, T.~J.~H. Vlugt, S.~Kjelstrup,
  D.~Bedeaux, {K}irkwood–{B}uff integrals: From fluctuations in finite
  volumes to the thermodynamic limit, The Journal of Chemical Physics 157~(13)
  (2022) 130901.
\newblock \href {https://doi.org/10.1063/5.0106162}
  {\path{doi:10.1063/5.0106162}}.

\bibitem{RN2861}
H.~Reiss, H.~L. Frisch, E.~Helfand, J.~L. Lebowitz, Aspects of the statistical
  thermodynamics of real fluids, The Journal of Chemical Physics 32~(1) (1960)
  119--124.
\newblock \href {https://doi.org/10.1063/1.1700883}
  {\path{doi:10.1063/1.1700883}}.

\bibitem{RN2871}
J.~P.~M. Postma, H.~J.~C. Berendsen, J.~R. Haak, Thermodynamics of cavity
  formation in water. {A} molecular dynamics study, Faraday Symposia of the
  Chemical Society 17 (1982) 55--67.
\newblock \href {https://doi.org/10.1039/FS9821700055}
  {\path{doi:10.1039/FS9821700055}}.

\bibitem{RN2983}
A.~Pohorille, L.~R. Pratt, Cavities in molecular liquids and the theory of
  hydrophobic solubilities, Journal of the American Chemical Society 112~(13)
  (1990) 5066--5074.
\newblock \href {https://doi.org/10.1021/ja00169a011}
  {\path{doi:10.1021/ja00169a011}}.

\bibitem{RN2997}
A.~Arvengas, E.~Herbert, S.~Cersoy, K.~Davitt, F.~Caupin, Cavitation in heavy
  water and other liquids, The Journal of Physical Chemistry B 115~(48) (2011)
  14240--14245.
\newblock \href {https://doi.org/10.1021/jp2050977}
  {\path{doi:10.1021/jp2050977}}.

\bibitem{RN2760}
V.~G. Baidakov, K.~S. Bobrov, Spontaneous cavitation in a {L}ennard-{J}ones
  liquid at negative pressures, The Journal of Chemical Physics 140~(18) (2014)
  184506.
\newblock \href {https://doi.org/10.1063/1.4874644}
  {\path{doi:10.1063/1.4874644}}.

\bibitem{RN2758}
V.~G. Baidakov, K.~R. Protsenko, Molecular dynamics simulation of cavitation in
  a {L}ennard-{J}ones liquid at negative pressures, Chemical Physics Letters
  760 (2020) 138030.
\newblock \href {https://doi.org/10.1016/j.cplett.2020.138030}
  {\path{doi:10.1016/j.cplett.2020.138030}}.

\bibitem{RN2715}
V.~Doebele, A.~Benoit-Gonin, F.~Souris, L.~Cagnon, P.~Spathis, P.-E. Wolf,
  A.~Grosman, M.~Bossert, I.~Trimaille, E.~Rolley, Direct observation of
  homogeneous cavitation in nanopores, Physical Review Letters 125~(25) (2020)
  255701.
\newblock \href {https://doi.org/10.1103/PhysRevLett.125.255701}
  {\path{doi:10.1103/PhysRevLett.125.255701}}.

\bibitem{RN2845}
J.~Puibasset, Cavitation in heterogeneous nanopores: The chemical ink-bottle,
  AIP Advances 11~(9) (2021) 095311.
\newblock \href {https://doi.org/10.1063/5.0065166}
  {\path{doi:10.1063/5.0065166}}.

\bibitem{RN2972}
M.~Bossert, A.~Grosman, I.~Trimaille, F.~Souris, V.~Doebele, A.~Benoit-Gonin,
  L.~Cagnon, P.~Spathis, P.-E. Wolf, E.~Rolley, Evaporation process in porous
  silicon: Cavitation vs pore blocking, Langmuir 37~(49) (2021) 14419--14428.
\newblock \href {https://doi.org/10.1021/acs.langmuir.1c02397}
  {\path{doi:10.1021/acs.langmuir.1c02397}}.

\bibitem{RN2996}
M.~Bossert, I.~Trimaille, L.~Cagnon, B.~Chabaud, C.~Gueneau, P.~Spathis, P.~E.
  Wolf, E.~Rolley, Surface tension of cavitation bubbles, Proceedings of the
  National Academy of Sciences of the United States of America 120~(15) (2023)
  e2300499120.
\newblock \href {https://doi.org/10.1073/pnas.2300499120}
  {\path{doi:10.1073/pnas.2300499120}}.

\bibitem{RN2991}
K.~M. Bal, E.~C. Neyts, Extending and validating bubble nucleation rate
  predictions in a {L}ennard-{J}ones fluid with enhanced sampling methods and
  transition state theory, The Journal of Chemical Physics 157~(18) (2022)
  184113.
\newblock \href {https://doi.org/10.1063/5.0120136}
  {\path{doi:10.1063/5.0120136}}.

\bibitem{RN3000}
K.~R. Protsenko, V.~G. Baidakov, Classical nucleation theory and molecular
  dynamics simulation: {C}avitation, Physics of Fluids 35~(1) (2023) 014111.
\newblock \href {https://doi.org/10.1063/5.0134778}
  {\path{doi:10.1063/5.0134778}}.

\bibitem{RN2989}
C.~P. Lamas, E.~Sanz, C.~Vega, E.~G. Noya, Estimation of bubble cavitation
  rates in a symmetrical {L}ennard-{J}ones mixture by {NVT} seeding
  simulations, The Journal of Chemical Physics 158~(12) (2023) 124109.
\newblock \href {https://doi.org/10.1063/5.0142109}
  {\path{doi:10.1063/5.0142109}}.

\bibitem{RN548}
J.~C. Fisher, The fracture of liquids, Journal of Applied Physics 19~(11)
  (1948) 1062--1067.
\newblock \href {https://doi.org/10.1063/1.1698012}
  {\path{doi:10.1063/1.1698012}}.

\bibitem{RN512}
M.~Blander, J.~L. Katz, Bubble nucleation in liquids, American Institute of
  Chemical Engineers Journal 21~(5) (1975) 833--848.
\newblock \href {https://doi.org/10.1002/aic.690210502}
  {\path{doi:10.1002/aic.690210502}}.

\bibitem{RN2920}
P.~G. Debenedetti, Metastable Liquids: Concepts and Principles, Princeton
  University Press, Princeton, NJ, 1996.

\bibitem{RN2913}
H.~Reiss, R.~K. Bowles, Some fundamental statistical mechanical relations
  concerning physical clusters of interest to nucleation theory, The Journal of
  Chemical Physics 111~(16) (1999) 7501--7504.
\newblock \href {https://doi.org/10.1063/1.480075}
  {\path{doi:10.1063/1.480075}}.

\bibitem{RN2756}
T.~Kinjo, M.~Matsumoto, Cavitation processes and negative pressure, Fluid Phase
  Equilibria 144~(1) (1998) 343--350.
\newblock \href {https://doi.org/10.1016/S0378-3812(97)00278-1}
  {\path{doi:10.1016/S0378-3812(97)00278-1}}.

\bibitem{RN2968}
V.~K. Shen, P.~G. Debenedetti, A computational study of homogeneous
  liquid–vapor nucleation in the {L}ennard-{J}ones fluid, The Journal of
  Chemical Physics 111~(8) (1999) 3581--3589.
\newblock \href {https://doi.org/10.1063/1.479639}
  {\path{doi:10.1063/1.479639}}.

\bibitem{RN2985}
A.~Vishnyakov, P.~G. Debenedetti, A.~V. Neimark, Statistical geometry of
  cavities in a metastable confined fluid, Physical Review E 62~(1) (2000)
  538--544.
\newblock \href {https://doi.org/10.1103/PhysRevE.62.538}
  {\path{doi:10.1103/PhysRevE.62.538}}.

\bibitem{RN2770}
Y.~W. Wu, C.~Pan, A molecular dynamics simulation of bubble nucleation in
  homogeneous liquid under heating with constant mean negative pressure,
  Microscale Thermophysical Engineering 7~(2) (2003) 137--151.
\newblock \href {https://doi.org/10.1080/10893950390203323}
  {\path{doi:10.1080/10893950390203323}}.

\bibitem{RN598}
A.~V. Neimark, A.~Vishnyakov, The birth of a bubble: a molecular simulation
  study, The Journal of Chemical Physics 122~(5) (2005) 054707.
\newblock \href {https://doi.org/10.1063/1.1829040}
  {\path{doi:10.1063/1.1829040}}.

\bibitem{RN2903}
J.~Wedekind, G.~Chkonia, J.~Wölk, R.~Strey, D.~Reguera, Crossover from
  nucleation to spinodal decomposition in a condensing vapor, The Journal of
  Chemical Physics 131~(11) (2009) 114506.
\newblock \href {https://doi.org/10.1063/1.3204448}
  {\path{doi:10.1063/1.3204448}}.

\bibitem{RN2766}
H.~Watanabe, M.~Suzuki, N.~Ito, Cumulative distribution functions associated
  with bubble-nucleation processes in cavitation, Physical Review E 82~(5)
  (2010) 051604.
\newblock \href {https://doi.org/10.1103/PhysRevE.82.051604}
  {\path{doi:10.1103/PhysRevE.82.051604}}.

\bibitem{RN2797}
S.~L. Meadley, F.~A. Escobedo, Thermodynamics and kinetics of bubble
  nucleation: Simulation methodology, The Journal of Chemical Physics 137~(7)
  (2012) 074109.
\newblock \href {https://doi.org/10.1063/1.4745082}
  {\path{doi:10.1063/1.4745082}}.

\bibitem{RN2901}
P.-R. ten Wolde, M.~J. Ruiz-Montero, D.~Frenkel, Simulation of homogeneous
  crystal nucleation close to coexistence, Faraday Discussions 104~(0) (1996)
  93--110.
\newblock \href {https://doi.org/10.1039/FD9960400093}
  {\path{doi:10.1039/FD9960400093}}.

\bibitem{RN2909}
P.~R. ten Wolde, D.~Frenkel, Enhancement of protein crystal nucleation by
  critical density fluctuations, Science 277~(5334) (1997) 1975--1978.
\newblock \href {https://doi.org/10.1126/science.277.5334.1975}
  {\path{doi:10.1126/science.277.5334.1975}}.

\bibitem{RN1956}
P.~R. ten Wolde, D.~Frenkel, Computer simulation study of gas-liquid nucleation
  in a {L}ennard-{J}ones system, The Journal of Chemical Physics 109~(22)
  (1998) 9901--9918.
\newblock \href {https://doi.org/10.1063/1.477658}
  {\path{doi:10.1063/1.477658}}.

\bibitem{RN2911}
S.~Auer, D.~Frenkel, Numerical prediction of absolute crystallization rates in
  hard-sphere colloids, The Journal of Chemical Physics 120~(6) (2004)
  3015--3029.
\newblock \href {https://doi.org/10.1063/1.1638740}
  {\path{doi:10.1063/1.1638740}}.

\bibitem{RN2905}
I.~Saika-Voivod, P.~H. Poole, R.~K. Bowles, Test of classical nucleation theory
  on deeply supercooled high-pressure simulated silica, The Journal of Chemical
  Physics 124~(22) (2006) 224709.
\newblock \href {https://doi.org/10.1063/1.2203631}
  {\path{doi:10.1063/1.2203631}}.

\bibitem{RN2893}
L.~Maibaum, Comment on ``elucidating the mechanism of nucleation near the
  gas-liquid spinodal'', Physical Review Letters 101~(1) (2008) 019601.
\newblock \href {https://doi.org/10.1103/PhysRevLett.101.019601}
  {\path{doi:10.1103/PhysRevLett.101.019601}}.

\bibitem{RN2895}
S.~Chakrabarty, M.~Santra, B.~Bagchi, {C}hakrabarty, {S}antra, and {B}agchi
  {R}eply, Physical Review Letters 101~(1) (2008) 019602.
\newblock \href {https://doi.org/10.1103/PhysRevLett.101.019602}
  {\path{doi:10.1103/PhysRevLett.101.019602}}.

\bibitem{RN2907}
S.~E.~M. Lundrigan, I.~Saika-Voivod, Test of classical nucleation theory and
  mean first-passage time formalism on crystallization in the {L}ennard-{J}ones
  liquid, The Journal of Chemical Physics 131~(10) (2009) 104503.
\newblock \href {https://doi.org/10.1063/1.3216867}
  {\path{doi:10.1063/1.3216867}}.

\bibitem{RN2847}
M.~A. González, G.~Menzl, J.~L. Aragones, P.~Geiger, F.~Caupin, J.~L.~F.
  Abascal, C.~Dellago, C.~Valeriani, Detecting vapour bubbles in simulations of
  metastable water, The Journal of Chemical Physics 141~(18) (2014) 18C511.
\newblock \href {https://doi.org/10.1063/1.4896216}
  {\path{doi:10.1063/1.4896216}}.

\bibitem{RN2801}
G.~Menzl, M.~A. Gonzalez, P.~Geiger, F.~Caupin, J.~L.~F. Abascal, C.~Valeriani,
  C.~Dellago, Molecular mechanism for cavitation in water under tension,
  Proceedings of the National Academy of Sciences USA 113~(48) (2016) 13582.
\newblock \href {https://doi.org/10.1073/pnas.1608421113}
  {\path{doi:10.1073/pnas.1608421113}}.

\bibitem{RN2954}
J.~Puibasset, A general relation between the largest nucleus and all nuclei
  distributions for free energy calculations, The Journal of Chemical Physics
  157~(19) (2022) 191102.
\newblock \href {https://doi.org/10.1063/5.0121580}
  {\path{doi:10.1063/5.0121580}}.

\bibitem{RN2966}
H.~M. Ellerby, H.~Reiss, Toward a molecular theory of vapor‐phase nucleation.
  {II}. {F}undamental treatment of the cluster distribution, The Journal of
  Chemical Physics 97~(8) (1992) 5766--5772.
\newblock \href {https://doi.org/10.1063/1.463760}
  {\path{doi:10.1063/1.463760}}.

\bibitem{RN154}
M.~P. Allen, D.~J. Tildesley, Computer Simulation of Liquids, Clarendon Press,
  Oxford, 1987.

\bibitem{RN51}
D.~Frenkel, B.~Smit, Understanding Molecular Simulation, Academic Press,
  London, 2002.

\bibitem{RN1330}
H.~Watanabe, N.~Ito, C.-K. Hu, Phase diagram and universality of the
  {L}ennard-{J}ones gas-liquid system, The Journal of Chemical Physics 136~(20)
  (2012) 204102.
\newblock \href {https://doi.org/10.1063/1.4720089}
  {\path{doi:10.1063/1.4720089}}.

\bibitem{RN1184}
Z.-J. Wang, C.~Valeriani, D.~Frenkel, Homogeneous bubble nucleation driven by
  local hot spots: {A} molecular dynamics study, The Journal of Physical
  Chemistry B 113~(12) (2009) 3776--3784.
\newblock \href {https://doi.org/10.1021/jp807727p}
  {\path{doi:10.1021/jp807727p}}.

\end{thebibliography}





\end{document}